# XUV exposed non-hydrostatic hydrogen-rich upper atmospheres of terrestrial planets. Part I: Atmospheric expansion and thermal escape

**Nikolai V. Erkaev[1,2], Helmut Lammer[3], Petra Odert[3,4], Yuri N. Kulikov[5], Kristina G. Kislyakova[3,4], Maxim L. Khodachenko[3], Manuel Güdel[6] Arnold Hanslmeier[4], Helfried Biernat[3]**

[1]Institute of Computational Modelling, Siberian Division of Russian Academy of Sciences, 660036 Krasnoyarsk, Russian Federation
(erkaev@icm.krasn.ru)
[2]Siberian Federal University, Krasnoyarsk, Russian Federation
[3]Austrian Academy of Sciences, Space Research Institute,
Schmiedlstr. 6, A-8042 Graz, Austria
(helmut.lammer@oeaw.ac.at, maxim.khodachenko@oeaw.ac.at,
kislyakova.kristina@oeaw.ac.at, petra.odert@oeaw.ac.at, helfried.biernat@oeaw.ac.at)
[4]Institute of Physics, University of Graz, Universitätsplatz 5, A-8010 Graz, Austria
(arnold.hanslmeier@uni-graz.at)
[5]Polar Geophysical Institute (PGI), Russian Academy of Sciences,
Khalturina Str. 15, Murmansk, 183010, Russian Federation
(kulikov@pgi.ru)
[6]Institute of Astrophysics, University of Vienna, Austria
(manuel.guedel@univie.ac.at)

Running Title: XUV exposed hydrogen-rich upper atmospheres

Corresponding Authors:
erkaev@icm.krasn.ru
Institute of Computational Modelling,
Siberian Division of Russian Academy of Sciences,
Akademgorodok 28/44 660036 Krasnoyarsk,
Russian Federation

**Submitted to ASTROBIOLOGY**

**ABSTRACT**

The recently discovered low-density "super-Earths" Kepler-11b, Kepler-11f, Kepler-11d, Kepler-11e, and planets such as GJ 1214b represent most likely planets which are surrounded by dense H/He envelopes or contain deep $H_2O$ oceans also surrounded by dense hydrogen envelopes. Although these "super-Earths" are orbiting relatively close to their host stars, they have not lost their captured nebula-based hydrogen-rich or degassed volatile-rich steam protoatmospheres. Thus it is interesting to estimate the maximum possible amount of atmospheric hydrogen loss from a terrestrial planet orbiting within the habitable zone of late main sequence host stars. For studying the thermosphere structure and escape we apply a 1-D hydrodynamic upper atmosphere model which solves the equations of mass, momentum and energy conservation for a planet with the mass and size of the Earth and for a "super-Earth" with a size of $2R_{Earth}$ and a mass of $10M_{Earth}$. We calculate volume heating rates by the stellar soft X-ray and EUV radiation and expansion of the upper atmosphere, its temperature, density and velocity structure and related thermal escape rates during planet's life time. Moreover, we investigate under which conditions both planets enter the blow-off escape regime and may therefore experience loss rates which are close to the energy-limited escape. Finally we discuss the results in the context of atmospheric evolution and implications for habitability of terrestrial planets in general.

**Key Words**: stellar activity, low mass stars, early atmospheres, Earth-like exoplanets, ENAs, ion escape, habitability

## 1. INTRODUCTION

The evolution of an Earth-like planet, where life on its surface may originate, is strongly related to its formation process, the impact history of the early planetary system, its initial water inventory, the escape of the early protoatmosphere and the host star-driven evolution of the remaining proto- or secondary atmosphere (e.g., Halliday, 2003; Lammer *et al.*, 2009; 2012a; Lammer, 2013). That the early atmospheres of terrestrial planets contained most likely more hydrogen than they do today, or were even hydrogen-dominated, was considered decades ago by researchers such as Holland (1962), Walker (1977), Ringwood (1979), Sekiya *et al.* (1980a; 1980b), Sekiya *et al.* (1981),Watson *et al.* (1981), and Ikoma and Genda (2006).

As illustrated in Fig. 1a, the earliest protoatmosphere of a planet should be a hydrogen-dominated H/He gas envelope which has been captured by the growing protoplanet from the system's nebula gas (e.g., Hayashi *et al.*, 1979). This is a complex process which depends on the formation time of the terrestrial planet, the nebula dissipation time, nebula opacity, the depletion factor of dust grains, the number and orbital parameters of additional planets, the protoplanet's gravity, its orbital location and the host star's radiation and plasma. Theoretical studies indicate that terrestrial planets may capture tens or even several hundreds of Earth ocean equivalent amounts of hydrogen around their rocky cores (e.g., Hayashi *et al.*, 1979; Mizuno, 1980; Wuchterl, 1983; Ikoma *et al.*, 2000; Ikoma and Gender, 2006; Rafikov, 2006).

Additionally to the captured nebula-based hydrogen envelopes, catastrophically outgassed steam atmospheres that depend on the impact history and the initial volatile content of a planet's interior can also be formed after accretion ends (see Fig. 1b) (Elkins Tanton and Seager, 2008; Elkins Tanton, 2011; Hamano *et al.*, 2013). Fig. 1b illustrates scenarios where primitive material with added $H_2O$ accreates into a planetary body. The additional water may be sufficient to oxidize all iron in the end member. The second scenario in Fig. 1b illustrates differentiated material where $H_2O$ and volatiles are added during the magma ocean phase. The resulting steam atmospheres depend then on the amount of volatiles which have been delivered or integrated during the growth or magma ocean phase of the planetary bodies (e.g., Elkins-Tanton, 2008).

Elkins-Tanton and Seager (2008) used bulk compositions corresponding to primitive and differentiated meteorite compositions and found that outgassing alone can create a wide range of planetary atmosphere masses which range from $\leq$ 1% of the planet's total mass up to ~6 % by mass of hydrogen, ~ 20 mass% of $H_2O$, and/or ~ 5 mass% of C compounds. According to their study hydrogen-rich atmospheres can also be outgassed as a result of oxidizing metallic Fe with $H_2O$. Depending on the initial volatile inventory and the depth of the magma ocean, during its solidification dense steam atmospheres with surface pressures ranging from ~$10^2$ - $10^4$ bar can be catastrophically outgassed (e.g., Bauer 1978; Abe 1993; 1997; Solomatov, 2000; Elkins-Tanton, 2003; 2008; Elkins-Tanton and Seager, 2008; Elkins-Tanton, 2011; Lammer, 2013).

In the case of the Solar System planets it is also expected that the early accretion stage, which resulted in the formation of planetesimals, most likely occurred in a highly reduced environment that resulted in large iron cores and volatiles of the planetesimals which were later delivered to protoplanets (e.g., Ringwood, 1979; Hunten *et al*., 1987; Dreibus *et al*., 1997). Wänke and Dreibus (1988) suggested that $H_2O$ would react with metallic iron in the accreting material to produce FeO or $Fe_2O_3$ by releasing H into the atmosphere. A more recent study by Rubie *et al.* (2009) found that during such high pressure phases, however, iron moves into a metallic state, preferentially to the oxidized phase, leaving $H_2O$ in the magma ocean.

Although, different terrestrial planets may accrete from differentiated planetesimals that contain different $H_2O$ and $CO_2$ contents, both molecules will enter solidifying minerals in small quantities (e.g., Elkins-Tanton, 2008; Elkins-Tanton and Seager, 2008). Elkins-Tanton (2008) showed that an Earth-sized terrestrial planet with a 2000 km deep magma ocean and an initial $H_2O$ and $CO_2$ content of ~ 0.05 wt. %, and ~ 0.01 wt. %, respectively, could build up a steam atmosphere of ~ 250 – 300 bar surface pressure. Higher initial $H_2O$ and $CO_2$ contents in the magma ocean of about ~ 0.5 wt. %, and ~ 0.1 wt. %, can result in water-dominated steam atmospheres of ≥ 3000 bar surface pressure (Elkins-Tanton, 2008; 2011). For an outgassed 300 bar steam atmosphere at early Earth, surface temperatures near the super-critical point of $H_2O$ could have been be reached within several to tens of Myrs after the formation of a magma ocean (Elins-Tanton, 2008; Hamano *et al*., 2013), while for larger "super-Earths" this timescale may be up to a factor 10 higher (Elkins-Tanton, 2011). After the surface of a young planet cools below the critical point, at an Earth-type planet with a surface pressure of ~ 220 bars corresponding to ~ 650 K, the supercritical fluid and steam atmosphere collapses into a liquid surface ocean (Elkins-Tanton, 2011; Lammer *et al.*, 2012a; Lammer, 2012). It is also possible that huge water oceans and steam atmospheres form directly from progressive solidification of a magma ocean if the initial $H_2O$ inventory is ≥ 1 wt % or closer to ~ 3 wt % which may be possible for "super-Earths" (Elkins-Tanton, 2011).

These theoretical findings have become very relevant since the discovery of several exoplanets that fall hypothetically within the rocky planet domain, such as the "super-Earths" Gliese 876d ($7M_{\text{Earth}}$) (Rivera *et al*., 2005), OGLE-2005-BLG-390Lb ($5$ $M_{\text{Earth}}$) (Gould et al., 2006), HD 69830 b ($10M_{\text{Earth}}$) (Lovis *et al*., 2006), Gliese 581c ($5.6M_{\text{Earth}}$), Gliese 581d ($\sim 8M_{\text{Earth}}$) (Beust *et al*., 2008) or Kepler-22b ($\sim 2.38R_{\text{Earth}}$) (Borucki *et al*., 2011). However, by knowing only the mass or the size, it is not possible to characterize a planet as rock-dominated, or as a mixed gaseous-rocky, or water-rocky-type planet. Fortunately, we know the size and mass of some "super-Earths" such as GJ 1214b ($R_{\text{pl}}=2.678R_{\text{Earth}}$; $M=6.55M_{\text{Earth}}$), 55 Cnc e ($R_{\text{pl}}=2R_{\text{Earth}}$; $M=8.63M_{\text{Earth}}$), CoRot-7b ($R_{\text{pl}}=1.58R_{\text{Earth}}$; $M=7.42M_{\text{Earth}}$), Kepler-10b ($R_{\text{pl}}=1.4R_{\text{Earth}}$; $M=4.56M_{\text{Earth}}$), Kepler-11b ($R_{\text{pl}}=1.97R_{\text{Earth}}$; $M=4.3M_{\text{Earth}}$), and Kepler-11f ($R_{\text{pl}}=2.61R_{\text{Earth}}$; $M=2.3M_{\text{Earth}}$) and the hypothesis that all terrestrial exoplanets which are expected to be rocky can lose their hydrogen or water-rich protoatmospheres, becomes testable (Lammer *et al.*, 2011; Lissauer et al., 2011; Lammer *et al*., 2012a; Lammer, 2013). Besides very close-in rocky "super-Earths", such as CoRoT-7b and Kepler-10b, that orbit their solar like host stars at about 0.0172 and 0.01684 AU, where both planets most likely lost their protoatmospheres completely (e.g., Leitzinger *et al.*, 2011), the other "super-Earths" are in more distant orbital locations and have lower mean densities, which suggests a presence of substantial envelopes of light gases, such as H/He or $H/H_2O$.

For instance GJ 1214b is a transiting "super-Earth" around an M star and has a mean density of $\sim 1.9$ g cm$^{-3}$, which is much lower than Earth's average density of $\sim 5.5$ g cm$^{-3}$. Because water has a density of about 1 g cm$^{-3}$, the chemical composition of GJ 1214b is most likely a mixture of rocks and $H_2O$ in liquid and gaseous form surrounded by an envelope of hydrogen (e.g., Charbonneau *et al*., 2009; Nettelmann *et al.*, 2011). This interesting composition is now confirmed by several independent studies (e.g., Miller-Ricci and Fortney, 2010; Croll *et al*., 2011; Nettelmann *et al.,* 2011). Recently Ikoma and Hori (2012) studied the low density "super-Earths" which populate the Kepler-11 system. The results of these authors indicate that indeed a huge amount of hydrogen may have been accumulated around these low-density "super-Earths" if the planetary disk dissipated slowly or the planets originated in cool environments. For explaining the observed densities of these "super-Earths" within a mass range between

~2 - 10$M_{Earth}$ they obtained captured hydrogen envelopes of several hundreds to thousands of Earth ocean equivalents (1EO$_{H2O}$ ~ $1.37 \times 10^{24}$ g → 1 EO$_H$ ~ $1.5 \times 10^{23}$ g) if one assumes a ~1% mass ratio of atmosphere to the total mass, and up to several $10^4$ EO$_H$ equivalents if one assumes a 10 % ratio. In a recent study by Lammer *et al.* (2013) that investigated the blow-off criteria of hydrogen-rich "super-Earths", including GJ 1214b, it was shown that GJ 1214b would experience hydrogen loss rates in the order of about $2.25 \times 10^{32} - 4 \times 10^{32}$ s$^{-1}$, that are too weak to remove its assumed hydrogen envelope or ocean during the planets remaining lifetime. Thus, planets such as GJ 1214b, or the low density "super-Earths" of the Kepler-11 system indicate that there may be many planets out there, which had most likely a different origin compared to Venus or Earth. These planets are representative for objects which are surrounded by a hydrogen-rich envelope of remaining gas from the protoplanetary nebula or dissociated $H_2O$ vapor.

From our brief discussion we conclude that terrestrial planetary atmospheres most likely initially originate with more or less dense hydrogen-dominated gaseous envelopes. Because present Venus or Earth are not surrounded by dense hydrogen envelopes and hydrogen is a light atom, it is generally assumed that these protoatmospheres escaped easily from early rocky Solar System planets, so that tectonic-related secondary outgassed atmospheres dominated the evolutionary process during later stages. However, as the discovered low-density "super-Earths" indicate that some planets may have a problem to get rid of their early protoatmospheres, it is important to understand if this is related to the planet's gravity, the stellar soft X-ray and extreme ultraviolet (XUV) flux, the stellar wind plasma flow, or if it depends on the planets growth time and related initial amount of captured or outgassed volatiles.

Therefore, the aim of the present work is to investigate how such hydrogen-rich protoatmospheres responded to the high XUV flux of the young and active host star if Earth-type planets orbit within the habitable zones of the main sequence stars. We model the structure of a hydrogen-dominated upper atmosphere and the related XUV-driven thermal escape rate from a planet with the size and mass of the Earth ($R_{pl}$=1$R_{Earth}$; $M_{pl}$=1 $M_{Earth}$) and from a "super-Earth" with the size of $R_{pl}$=2$R_{Earth}$ and a mass $M_{pl}$=10$M_{Earth}$ when their thermospheres were exposed to XUV fluxes from 1 to 100 times that of today's Sun, which is the range typical for planets inside the habitable zone (e.g., Ribas *et* *al.*, 2005; Scalo *et al.*, 2007; Claire *et al.*, 2012). We describe in Sect. 2 the applied hydrodynamic upper atmosphere model. In Sect. 3 we present density, velocity and temperature profiles as well as expected thermal atmospheric escape rates as a function of the stellar XUV flux. We compare our results with previous studies related to a hydrogen-rich early Earth and discuss in Sect. 4 the implications of our results to habitability aspects of terrestrial planets in general. Finally, this work presents the basic input parameters for Part II of these investigations (Kislyakova *et al.*, 2013), in which we study the stellar wind plasma interaction with a hydrogen-rich planetary upper atmosphere within the habitable zone of a M-type host star, the formation of extended planetary hydrogen coronae and the stellar wind plasma-induced non-thermal ion pick up escape process from similar hydrogen-rich test planets.

## 2. XUV-INDUCED HYDRODYNAMIC UPPER ATMOSPHERE EXPANSION AND ESCAPE

Depending on the host star's energy input into the upper atmosphere of a planet, one can split up the atmospheric escape processes in two categories: thermal and non-thermal escape.

Thermal atmospheric escape can again be divided into the classic Jeans escape and hydrodynamic outflow which can result in blow-off (e.g., Chamberlain, 1963; Öpik 1963; Bauer and Lammer 2004; *Tian et al.*, 2008a; Tian *et al.*, 2008b; Lammer, 2013). In the first case atmospheric particles, that populate the high-energy tail of a Maxwell distribution at the exobase level, where the mean free path $l$ equals the scale height $H$

$$l \approx H = \frac{kT_{exo}}{mg_{exo}}, \tag{1}$$

are lost from the exosphere of a planet. Here $k$ is the Boltzmann constant, $T_{\text{exo}}$ and $g_{\text{exo}}$ the temperature and gravitational acceleration at the exobase level $r_{\text{exo}}$, and $m$ is the mass of the main atmospheric species at the exobase distance. Under high XUV flux conditions the thermosphere starts to expand dynamically accompanied by adiabatic cooling so that the exobase location reaches several planetary radii (Watson *et al.*, 1981; Tian *et al.*, 2005a; Lammer *et al.*, 2008; Tian *et al.*, 2008a; 2008b; Lammer *et al.*, 2012a; Lammer, 2013).

In this second case, which is relevant for the present work, the upper atmosphere is not hydrostatic anymore so that hydrodynamic blow-off at the exobase level may occur, which results in the evaporation of the whole exosphere. However, under certain XUV conditions and planetary parameters the thermospherere can hydrodynamically expand but not all atoms may reach escape velocity and the loss results in a strong Jeans-type escape rate from an expanded exobase level.

## *2.1 Atomic vs. molecular hydrogen: breakdown of $H_2$ molecules*

Because we are interested to know how the nebula-based hydrogen envelopes or outgassed steam atmospheres respond to and escape thermally due to the XUV flux of a young star, it is important to know if hydrogen dominates the upper atmosphere in molecular or atomic form. In a previous study related to the dynamics of escaping hydrogen-dominated upper atmospheres from early Earth, Tian *et al.* (2005a) assumed that $H_2$ is the main thermospheric species in the upper atmosphere. However, that study focused on ~ 5 times higher XUV fluxes compared to today's solar value. Such values are expected for the young Sun about ~3.5 Gyr ago when life originated on Earth. According to astrophysical data gathered by multi-wavelength satellite observations of solar proxies with younger ages, it is known that the XUV flux of a young solar like star is saturated at ~100 times of the average present time solar value during the first 100 Myr (e.g., Güdel *et al.,* 1997; Ribas *et al.,* 2005; Güdel 2007) and decreases during the first Gyr following a power law to a value which yields XUV flux enhancement values which are ~10 times higher compared to today's Sun ~4 Gyr ago and ~5 times ~3.5 Gyr ago (Ribas *et al*., 2005; Claire *et al.,* 2012). For lower mass M-type stars this saturation can last longer before the flux of the short wavelength radiation decreases according to the similar power law as for the solar like stars (Scalo *et al*., 2007). Therefore, at the time when the nebula-based or outgassed hydrogen-rich protoatmospheres originate, the XUV flux values are much higher than applied in the study of the early Earth by Tian *et al.* (2005a). From photochemical studies of hydrogen-dominated "Hot Jupiter" thermospheres it is known that $H_2$ molecules break down to H atoms if the XUV flux is > 25 times that of the present solar value (Yelle, 2004; Koskinen *et al.*, 2010). The $H_2$ molecules will be dissociated via reaction

$$H_2 + h\nu \rightarrow H + H. \tag{2}$$

In case $H_2^*$ is a vibrationally excited hydrogen molecule H atoms can also be produced via reaction

$$H^+ + H_2^* \rightarrow H_2^+ + H. \tag{3}$$

For atmospheric temperatures which are $\geq 2000$ K, $H_2$ molecules break thermally into atomic hydrogen K (Koskinen *et al.*, 2010)

$$H_2\left(T_{dis} \geq 2000K\right) \rightarrow H + H. \tag{4}$$

For steam atmospheres one can also expect that atomic hydrogen dominates the upper atmosphere because the XUV flux dissociates the water molecules in H and OH. Even on present Earth with present time XUV radiation the $H_2$ molecule number density is one order of magnitude lower compared to that of H atoms above 100 km altitude. According to Koskinen *et al.* (2010) for XUV fluxes > 25 times of that of the present Sun one can expect a stronger decrease of $H_2$ in hydrogen dominated thermospheres and therefore a higher amount of H atoms in the upper atmosphere. Model simulations of "hot Jupiters" which are exposed to 450 times higher XUV fluxes compared to that of the present solar value indicate that a large fraction of the neutral hydrogen atoms are ionized at distances $\geq 3$ $R_{pl}$ (Yelle, 2004; Grazia-Muñoz, 2007; Koskinen *et al.*, 2012). In case of the hydrogen-rich gas giant HD 209458b this distance corresponds to ~30 Earth-radii. Because our test planets are exposed to more than 4.5 times lower XUV flux values compared to that of typical "hot Jupiters" it is justified to assume that neutral H atoms dominate over the ionized component up to the exobase level.

Additional possibilities for the production of atomic hydrogen are photochemical reactions between $CH_4$ and other hydrocarbons below and close to the homopause level (e.g., Atreya, 1986; Atreya, 1999). Because of these processes atomic hydrogen populates the upper atmosphere above the homopause level of Jupiter, Saturn, Uranus and Neptune. Since the studied hydrogen-rich test-planets are much hotter compared to the hydrogen-rich Solar System gas and ice giants and are exposed to higher photon fluxes and may also contain hydrocarbons in their lower atmospheres (Kuchner, 2003), we assume in this particular investigation that atomic hydrogen is most likely the dominant species above the homopause levels of the studied planets.

However, for low XUV fluxes or if one assumes that IR-cooling molecules, such as $H_3^+$, $CO_2$, etc. may decrease the thermospheric temperature and, hence, decrease thermal dissociation of $H_2$, it is instructive to compare the modeled upper atmosphere structure calculated for pure atomic and molecular hydrogen atmospheres. Therefore, for comparing the difference of a $H_2$ dominated thermosphere with an H dominated, we model both scenarios.

We should also note that it was shown in Lammer *et al.* (2011) and Lammer (2013), if an outgassed steam atmosphere results in very high surface pressures such as several ~ $10^3$ - $10^4$ bar (Elkins-Tanton, 2011) than, the planet might have a problem in losing the remaining oxygen in a similar amount compared to the hydrogen. This result agrees with the suggestion of Kasting (1995) and Chassefière (1996a; 1996b) that there may be planets, depending on their size, mass, orbital distance, as well as their host star's XUV flux evolution, which could accumulate huge amount of abiotic oxygen. In this study we focus only on the escape of hydrogen, either envelopes which remained from captured nebula gas or the part which is lost from a steam atmosphere. In the future we plan also to study the accompanied thermal and non-thermal escape of oxygen from steam atmospheres.

*2.2 Model description and numerical scheme*

For studying the response of hydrogen-dominated upper atmospheres of an Earth-like and a "super-Earth"-type planet within the habitable zone to the stellar XUV flux we apply a time-dependent 1-D hydrodynamic upper atmosphere model, which solves the system of the fluid equations for mass, momentum, and energy conservation in spherical coordinates

$$\frac{\partial \rho r^2}{\partial t} + \frac{\partial \rho v r^2}{\partial r} = 0,$$

$$\frac{\partial \rho v r^2}{\partial t} + \frac{\partial r^2 (\rho v^2 + P)}{\partial r} = -\rho g r^2 + 2Pr,$$

$$\frac{\partial r^2 [\, \rho v^2/2 + P/(\gamma - 1)\,]}{\partial t} + \frac{\partial v r^2 (\, \rho v^2/2 + \gamma P/(\gamma - 1))}{\partial r} = -\rho v r^2 g + q\, r^2, \qquad (6)$$

$$P = \frac{\rho}{m} kT, \quad g = G \frac{M_p}{r^2}.$$

Here $r$ is the radial distance from the center of the planet, $\rho$, $v$, $P$, $T$ are the mass density, radial velocity, pressure and temperature, respectively, $m$ is the mass of particle, $k$ is the Boltzmann constant, $G$ is the gravitational constant, and $\gamma$ is the polytropic index which corresponds to the ratio of the specific heats. For computational convenience we introduce the following normalizations

$$\tilde{P} = P/(n_0 k T_0), \quad \tilde{\rho} = \rho/(n_0 m), \quad \tilde{v} = v/v_0, \quad v_0 = \sqrt{kT_0/m}, \quad \tilde{T} = T/T_0,$$
$$\beta_0 = GmM_{pl}/(R_0 k T_0), \quad \tilde{q} = qR_0/(mn_0 v_0^3), \quad \tilde{r} = r/R_0, \quad \tilde{t} = tv_0/R_0. \tag{7}$$

Here $R_0$ is the altitude of the lower thermosphere (homopause) which is close to the planetary radius $R_{pl}$, $n_0, v_0, T_0$ are the density, thermal velocity and temperature at the lower boundary, which corresponds to the lower part of the thermosphere around the homopause distance. $\beta_0$ is the Jeans escape parameter at the lower boundary. This parameter is $\leq 1.5$ at the corresponding exobase level, if an upper atmosphere reaches blow-off conditions. After introducing these normalizations we obtain the following system of equations

$$\frac{\partial \tilde{\rho} \tilde{r}^2}{\partial \tilde{t}} + \frac{\partial \tilde{\rho} \tilde{v} \tilde{r}^2}{\partial \tilde{r}} = 0,$$
$$\frac{\partial \tilde{\rho} \tilde{v} \tilde{r}^2}{\partial \tilde{t}} + \frac{\partial \tilde{\rho} \tilde{r}^2 (\tilde{v}^2 + \tilde{T})}{\partial \tilde{r}} = -\tilde{\rho}(\beta_0 - 2\tilde{T}\tilde{r}), \tag{8}$$
$$\frac{\partial \tilde{\rho} \tilde{r}^2 (\tilde{v}^2/2 + \tilde{T}/(\gamma-1))}{\partial \tilde{t}} + \frac{\partial \tilde{\rho} \tilde{v} \tilde{r}^2 (\tilde{v}^2/2 + \gamma\tilde{T}/(\gamma-1))}{\partial \tilde{r}} = -\tilde{\rho} \tilde{v} \beta_0 + \tilde{q} r^2.$$

The thermal conductivity can be estimated as $c_v V_T/\sigma$. Therefore, the energy flux per one steradian (sr$^{-1}$) due to the thermal conduction can be estimated as

$$W_c = \frac{c_v V_0 R_0 T_0}{\sigma_c} \left( \tilde{r}^2 \sqrt{\tilde{T}} \frac{\partial \tilde{T}}{\partial \tilde{r}} \right), \tag{9}$$

where $\sigma_c$ is collisional cross section. If one compares this energy flux $W_c$ with the total energy flux given by the energy-limited formula (see Eq. 31) we find that the ratio of these energies is quite small

$$\frac{W_c}{\left( \Gamma_{th} G \dfrac{mM_{pl}}{R_{pl}} \right)} \approx 10^{-3} \ \tilde{r}^2 \sqrt{\tilde{T}} \ \frac{\partial \tilde{T}}{\partial \tilde{r}} << 1. \tag{10}$$

Fig 2 compares the thermal energy flux due to hydrodynamic flow per one steradian of the atmospheric particles with the convective thermal energy flux obtained by our hydrodynamic solutions for a H dominated upper atmosphere of an Earth-like planet which is exposed to a 10 times higher XUV flux compared to that of today's Sun. The two sudden decreases in the convective thermal energy flux can be explained because it is proportional to the gradient of the temperature, and therefore it decreases in the vicinity of the temperature maximum and minimum. At first point we have strong temperature maximum, and at the second point we have shallow temperature minimum.

Therefore, one can conclude that the influence of the thermal conduction on the atmospheric escape is expected to be rather small so that we can neglect the thermal conduction term in the energy equation (8). For the description of this numerical scheme we rewrite our system of equations in vector form

$$\frac{\partial U}{\partial t} + \frac{\partial \, \Gamma(U)}{\partial r} = Q(U), \tag{11}$$

with

$$U = \begin{pmatrix} \tilde{\rho}r^2 \\ \tilde{\rho}\tilde{v}\tilde{r}^2 \\ \tilde{\rho}\tilde{r}^2(\tilde{v}^2/2 + \tilde{T}/(\gamma-1)) \end{pmatrix} \quad \Gamma = \begin{pmatrix} \tilde{\rho}\tilde{v}\tilde{r}^2 \\ \rho\tilde{r}^2(\tilde{v}^2 + \tilde{T}) \\ \tilde{\rho}\tilde{v}\tilde{r}^2(\tilde{v}^2/2 + \gamma\tilde{T}/(\gamma-1)) \end{pmatrix} \quad Q = \begin{pmatrix} 0 \\ -\tilde{\rho}(\beta - 2\tilde{T}\tilde{r}) \\ -\tilde{\rho}\tilde{v}\beta + \tilde{q}r^2 \end{pmatrix} \tag{12}$$

and apply the finite difference numerical scheme of MacCormack (1969), which is of the second order of accuracy. In computational fluid dynamics, this method is widely used as a discretization scheme for the numerical solution of hyperbolic partial differential equations. After we introduce the MacCormack scheme, we obtain the following equations

$$U_i^{n+1/2} - U_i^n = \Delta t \left[ -\frac{(\Gamma_{i+1}^n - \Gamma_i^n)}{(r_{i+1} - r_i)} + Q_i^n \right],$$

$$U_i^{n+1} = 0.5(U_i^{n+1/2} + U_i^n) + \frac{\Delta t}{2} \left[ -\frac{(\Gamma_i^{n+1/2} - \Gamma_{i-1}^{n+1/2})}{(r_i - r_{i-1})} + Q_i^{n+1/2} \right]. \tag{13}$$

Here index "*i*" numerates grid points along the radial direction, and "*n*" is the number of a time step. We use a non-uniform grid: $r_i = (r_{max})^{i/n}$. The grid size is an increasing function of the radial distance *r*.

Similar to Tian *et al.* (2005a), we assume that the lower boundary is fixed at the base of the thermosphere, which corresponds to an altitude distance of ~100 km. At the lower boundary we set constant values for density and temperature and a "free" condition for velocity

$$\tilde{n}_1 = 1, \quad \tilde{T}_1 = 1, \quad \tilde{v}_1 = \tilde{v}_2, \quad \tilde{r} = 1. \tag{14}$$

The upper boundary is taken at about 20 $R_{pl}$ but the results of our hydrodynamic model are considered as accurate only up to the exobase level which separates the collision dominated thermosphere from the collision-less exosphere, because above this distance hydrodynamics is not valid. At the upper boundary we set free boundary conditions for the velocity, density and temperature

$$\tilde{v}_n = \tilde{v}_{n-1}, \quad \tilde{n}_n = \tilde{n}_{n-1}, \quad \tilde{T}_n = \tilde{T}_{n-1}, \quad \tilde{r} = \tilde{r}_1 = 20. \tag{15}$$

As an initial condition for density we assume the Boltzmann distribution

$$\tilde{n}(r) = \exp\left[\beta_0\left(-1 + \frac{1}{\tilde{r}}\right)\right]. \tag{16}$$

The initial temperature is assumed to be constant, $T = T_0$. For the velocity we introduce the initial flow

$$\tilde{v}(r) = 0.01(\tilde{r} - 1). \tag{17}$$

It is worth noting that the final steady state solution does not depend on the initial conditions. For tests of our numerical code we compare the calculated velocity profile with the analytical solution of Parker (1958). This comparison is shown in Fig. 3 for a thermal escape parameter $\beta_0$ of 5.

### 2.3 Energy absorption and heating function

The applied energy absorption model and its geometry, which is used for the establishment of a functional dependence between the XUV volume heating rate $q(r, \theta)$ and the distance *r* is illustrated in Fig. 4.

The stellar XUV flux outside the planet is illustrated in Fig. 4 by the arrows. The incoming XUV flux decreases due to absorption in the thermosphere, which results in dissociation and ionization and, hence, in heating of the upper atmosphere. The equation for the XUV flux transfer can be then be written as

$$\frac{dJ}{dz} = J\,\sigma_{XUV}\,n(\sqrt{z^2+d^2}),$$  (18)

where $\sigma_{xuv}$ is the XUV absorption cross section and $J$ is the XUV flux in units of erg cm$^{-2}$ s$^{-1}$. By integrating eq. (18) along $z$, we find

$$J(z,d) = J_0 \exp\left(-\int_z^\infty \sigma_{XUV}\, n(\sqrt{s^2+d^2}\,)ds\right),$$  (19)

where $J_0$ is the XUV flux at the planet's orbit but outside the atmosphere. In spherical coordinates eq. (19) can be written as

$$J(r,\theta) = J_0 \exp\left(-\int_{r\cos(\theta)}^\infty \sigma_{XUV}\, n(\sqrt{s^2+r^2\sin^2(\theta)}\,)ds\right).$$  (20)

The XUV volume heating rate $q(\mathrm{r},\theta)$ is determined by the following equation

$$q(r,\theta) = \eta\sigma_{XUV}\,n(r)\,J(r,\theta).$$  (21)

In case $\theta = 0$, Eqs. (20) and (21) are similar as those used by Murray-Clay et al. (2009). Here $\eta$ is the heating efficiency, that defines the percentage of incoming XUV energy which is transferred locally into heating of the gas. Eq. (21) can be applied everywhere besides the shadow zone shown in Fig. 4. By averaging the XUV volume heating rate over the planet's dayside we find

$$\overline{q}(r) = \eta\sigma_{XUV}\,n(r)\,\frac{1}{4\pi}\int_0^{\pi/2+\arccos(1/r)} J(r,\theta)\,2\pi\sin(\theta)d\theta.$$  (22)

Finally we obtain the normalized XUV volume heating rate in the following form

$$\widetilde{q}(r) = B\widetilde{n}(r)\,\frac{1}{2}\int_0^{\pi/2+\arccos(1/r)}\exp\left(-\int_{r\cos(\theta)}^\infty a\,\widetilde{n}(\sqrt{s^2+r^2\sin^2(\theta)}))\right)\sin(\theta)d\theta,$$  (23)

with the coefficient

$$B = \frac{\eta J_{\mathrm{XUV0}} \sigma_{XUV} R_0}{m \, v_0^3}, \quad a = \sigma_{XUV} n_0 R_0. \tag{24}$$

In the present parameter study we use the integrated XUV flux and do not consider a wavelength dependence of the incoming stellar XUV radiation. For understanding the effect of the spectral dependence on the volume heating rate the model will be extended in this direction for future studies. By integrating Eq. (23) over the whole domain we obtain the total energy absorption in normalized units, which is proportional to the incoming stellar XUV flux. The appropriate coefficient $B$ is chosen in order to satisfy Eq. (23) for a given value of $J_{xuv0}$. However, eq. (23) is rather complex we introduce the following approximation

$$\tilde{q}(r) = 0.767 B \tilde{n}(r) \exp\left(-\int_r^\infty 0.6 \mathrm{as}(s^{1/3}+1)\tilde{n}(s)ds\right), \tag{25}$$

which is more convenient and less time consuming for the numerical calculations. Fig. 5 shows the heating rate normalized to its maximum value $q_{\max}$ which can be written as

$$q_{\max} = 4.27 \times 10^{-5} B, \tag{26}$$

compared to the function given in eq. (25). One can see in Fig. 5 that both expressions yield nearly similar results and eq. (25) shows a negligible difference from the curve which results from eq. (21). Therefore, we apply the less complex formula given by eq. (25) for the numerical calculations of the hydrodynamic outflow regimes of the hydrogen-rich terrestrial test planets.

Apart from the expansion-related adiabatic cooling of the thermosphere, possible IR-cooling molecules such as $CO_2$, or $H_3^+$ may also modify the upper atmosphere structure and the related thermal escape rates. Because it is not known how many IR-cooling species may be available in hydrogen-rich protoatmospheres, we investigate their possible effect on thermospheric cooling by introducing different heating efficiency values. Photochemical models indicate that the heating efficiency $\eta$ lies for hydrogen-rich atmospheres of "hot Jupiters" within the range of ~ 35 % - 60 % (Yelle, 2004; Lammer *et al.*, 2009; Koskinen *et al.*, 2013), but may be lower if more IR-cooling molecules which are more stable and abundant in not so extreme radiation environments are present in the lower thermosphere.

Because the presence of minor species or IR-cooling molecules in hydrogen-rich protoatmospheres of terrestrial exoplanets is unknown and a detailed photochemical study of hydrogen-rich of terrestrial planets is beyond the scope of this work, we study a possible influence of IR-cooling molecules in H and $H_2$ dominated upper atmospheres by applying two $\eta$ values, a lower value of 15 % and a higher one of 40 %, which agrees with the calculations of Koskinen *et al.* (2013) for the hydrogen-rich gas giant HD 209458b.

### 2.4 Initial conditions and input parameters

Tian *et al.* (2005a) modeled a hydrogen-rich upper atmosphere of early Earth for solar XUV flux vales which were expected during the Archean era (XUV ~ 5 times the present solar value). In the present parameter study we assume similar initial input parameters for the hydrogen density ($n_0 = 5 \times 10^{12}$ cm$^{-3}$) and "skin" or equilibrium temperature ($T_0 = 250$ K) and the adiabatic index $\gamma = 5/3$ at the homopause level.

We assume that the upper atmosphere of an Earth-like planet *($R_{pl}=1R_{Earth}$ and $M_{pl}=1M_{Earth}$)* and of a "super-Earth" *($R_{pl}=2R_{Earth}$ and $M_{pl} =10M_{Earth}$)* are located within the habitable zone at 1 AU of a solar-like G star or a corresponding orbital distance within K and M star habitable zones and are exposed to XUV flux values in a range between 1 and 100 times that of today's Sun. According to astrophysical observations of young solar proxies, one can expect that these flux values cover the whole XUV flux range for planets orbiting inside the habitable zone (e.g., Ribas *et al.*, 2005; Güdel, 2007; Scalo *et al.*, 2007; Claire *et al.*, 2012) of their host stars. The chosen lower boundary number density $n_0$ of $5 \times 10^{12}$ cm$^{-3}$ corresponds to an extremely high $H_2O$ mixing ratio $f_{H2O}$ at the mesopause or atmospheric cold trap of ~ 50 % but lies also close to the value of Neptune's homopause hydrogen density of $\sim 10^{13}$ cm$^{-3}$ which is a remnant of nebula accumulated gas (e.g., Atreya *et al.*, 1999). Kasting and Pollack (1983) showed that Greenhouse effects in a lower atmosphere can raise the $H_2O$ mixing ratio $f_{H2O}$ to values which are $\geq 5\times10^{-4}$ near the cold trap so that hydrogen which originates from the dissociation of $H_2O$ molecules will dominate the whole upper atmosphere. Because we do not investigate the hydrogen sources of our upper atmosphere, which can originate from the dissociation of hydrogen containing volatiles but also from remaining nebula gas, in detail we do not assume different hydrogen mixing ratios in this particular study. However, we note that lower hydrogen content calculations are very important for defining the inner boundary of the habitable zone.

Greenhouse effects near the surface could be related to remnants of nebula-based hydrogen envelopes, the available content of greenhouse gases ($CO_2$, $CH_4$, etc.) near the surface, or impacts (e.g., Ikoma and Genda, 2006; Pierrehumbert and Gaidos, 2011; Kopparapu *et al.*, 2013).

The number density $n_0$ at the base of the thermosphere can never be arbitrarily increased or decreased as much as by an order of magnitude from its real atmospheric value. In fact the density $n_0$ is a conservative value and cannot change much even if the surface pressure on a planet varies during its life time by many orders of magnitude. The reason for this is that the value of $n_0$ is strictly determined by the XUV absorption optical depth of the thermosphere above its base level, $r_0$. Because, of this, the base of the thermosphere is conventionally defined as a level $r_0$ in the upper atmosphere where the bulk of the incident solar XUV radiation averaged over $\lambda$ in the XUV absorption range, $\Delta\lambda_{XUV}$, and over zenith angles $\chi$ from 0 to $\pi/2$ is completely absorbed by the above lying layers of the atmosphere

$$\langle\tau_0\rangle = n_0\,H_0\,\langle\sigma_H\rangle\,\langle Ch_0\rangle = (P_0/m_H g_0)\,\langle\sigma_H\rangle\,\langle Ch_0\rangle \qquad (27)$$

where $\langle\tau_0\rangle$ is the optical depth of the thermosphere averaged over the XUV wavelength range $\Delta\lambda_{XUV}$ and solar zenith angle $\chi$ at which the bulk of the solar XUV radiation is completely absorbed, $\langle\sigma_H\rangle$ is the absorption cross-section of H averaged over $\Delta\lambda_{XUV}$; and $H_0$ and $\langle Ch_0\rangle$ are the scale height and a Chapman function at the base $r_0$. The atmospheric pressure $P_0$ at the thermosphere base $r_0$ for any dense atomic hydrogen dominated atmosphere with a surface pressure $P_s \geq P_0$, should then be a constant value which is defined by the following equation

$$P_0 = (\langle\tau_0\rangle\,m_H\,g_0)\,/\,(\langle Ch_0\rangle\,\langle\sigma_H\rangle) \qquad (28)$$

Because of the perfect gas law $P = n\,k\,T$, the base number density $n_0$ is inversely proportional to the base temperature $T_0$ so that the variation of the number density at the base of the thermosphere (and the density itself) will always have a limited range of values. These values are determined only by the variation of a skin temperature of a planet, to which the base temperature $T_0$ is usually quite close. In a denser atmosphere, the base of the thermosphere will simply rise to a higher altitude where the base pressure $P_0$ retains the same constant value as in a less dense atmosphere.

## 3 RESULTS

By assuming the before discussed atmospheric input parameters for the two hydrogen-rich test planets we model for both the volume heating rates, the upper atmosphere structures up to the exobase levels, the exobase temperatures and finally the thermal escape rates for solar XUV flux values which are similar that that of the present solar value in 1 AU and for 5, 10, 50 and 100 times higher fluxes compared to that of today's Sun.

### *3.2 Upper atmosphere heating, expansion and structure*

In the following section we present results related to the heating of the upper atmosphere, its expansion and related escape rates for the hydrogen-rich Earth-like planet and the "super-Earth" which are exposed by stellar XUV fluxes between 1 and 100 times that of today's Sun.

Table 1 show the exobase temperature $T_{exo}$, exobase distance $R_{exo}$ and the distance of the transonic point $R_s$ in planetary radii $R_{pl}$ of the Earth-like planet for a thermosphere which is dominated by hydrogen atoms and molecules. The heating efficiency $\eta$ is taken to be 15 % and 40 % for the H dominated upper atmosphere which is exposed to various stellar XUV flux values normalized to that of the present solar value in 1 AU. For the $H_2$ dominated upper atmosphere we the lower value $\eta$=15 % is chosen.

One can see that the exobase level for a molecular hydrogen dominated upper atmosphere expands less if compared to a thermosphere which is dominated by atomic hydrogen, while the exobase temperature for the less expanded $H_2$ dominated thermosphere is hotter. The reason for this behavior is, that the lighter H atoms experience stronger adiabatic cooling compared to the $H_2$ molecules. For a higher $\eta$ of 40 % the exobase level expands further outwards from the planetary surface and the exobase temperature $T_{exo}$ becomes hotter compared to the 15 % cases. Table 2 shows the similar parameters and modeled results for the more massive and larger "super-Earth". One can see that the adiabatic cooling and the related expansion of the upper atmosphere is weaker compared to the Earth-like test planet so that $T_{exo}$ remains hotter. Only for high

XUV flux values > 50 times that of today's Sun the XUV heating is stronger compared to the adiabatic cooling so that $T_{exo}$ of the Earth-like test planet is higher than that of the "super-Earth". One can also see in Table 2 that $T_{exo}$ for the "forced" $H_2$ dominated thermosphere with XUV fluxes $\geq$ 50 times compared to that of the present solar value are higher than the dissociation temperature of $H_2$ molecules. In these cases dissociated hydrogen atoms would expand the exobase level further outward so that it may lie between ~8 – 11 and ~10 – 13 $R_{pl}$. For the Earth-like H dominated planet and a heating efficiency $\eta$ of 40%, 2000 K is reached for XUV flux values which are $\geq$ 30 times of the present solar value and for the 100 XUV flux value the thermosphere is hotter when compared to the H-dominated more massive "super-Earth".

The main reason for this unexpected behavior is the following: If an Earth-like planet with its lower gravity is exposed to high XUV fluxes the number density of the upper atmosphere decreases much slower, and because the XUV volume heating rate is proportional to the atmospheric density the energy is absorbed over a wider range compared to the more massive "super-Earth". Therefore, the XUV heating overcomes the adiabatic cooling and yields for a higher exobase temperature for the lower mass Earth-like planet. One can see from both tables that for high XUV fluxes the exobase levels can also expand to distances which are $\geq$ 10 $R_{pl}$. In such cases one can expect that the "super-Earth" produces a large exosphere or atomic hydrogen corona which is most likely not protected by an intrinsic magnetosphere so that the exospheric neutral particles will interact with the stellar wind and can be lost by non-thermal ion escape processes such as ion pick up (Kislyakova *et al*., 2012).

One can see from Table 1 that for the H-rich Earth-like planet with a heating efficiency $\eta$=15 % the sonic transition point $R_s$ lies about one Earth-radii above the exobase level and below for higher XUV cases. For a similar planet with a higher $\eta$ value of 40 % $R_s$ is reached at the exobase level for XUV flux value > 7 time that of today's solar value. For an Earth-like planet with an upper atmosphere which is dominated by $H_2$ molecules blow-off is reached for XUV fluxes which are higher than 40 times that of the present Sun. In case of the hydrogen-rich "super-Earth", one can see that Rs lies above the exobase level for a heating efficiency $\eta$=15 % at 100 XUV and for values which are higher than ~ 40 times of the solar value for a higher $\eta$ of 40 %. In case of a $H_2$ dominated thermosphere the "super-Earth" does not reach blow-off conditions with $\eta$=15 % within the studied XUV range.

Fig. 6 compares the volume heating rate profiles up to the exobase level for 1 XUV (long dashed lines), 5 XUV (dashed-dotted-dotted lines), 10 XUV (dashed-dotted lines), 50 XUV (dashed-lines) and 100 XUV (dotted lines) cases between the Earth-like planet (Figs. 5a-c) and the more massive "super-Earth" (Figs. 6d-f). Fig. 6a and Fig. 6c correspond to profiles of H- and Fig. 6b and Fig. 6e to $H_2$-dominated upper atmospheres and heating efficiencies of 15 %. Figs. 6c and 6f belong to H-dominated upper atmospheres with 40 % heating efficiency.

Fig. 7 shows the corresponding temperature, density and velocity structures of the hydrogen-rich non-hydrostatic expanded upper atmosphere from the lower thermosphere up to the exobase level for a heating efficiency $\eta = 15$ % and XUV fluxes, which are 1, 5, 50 and 100 times higher compared to today's solar value for both H-rich test planets.

One can see that the upper atmosphere does not expand to the same large distances compared to a thermosphere which would be dominated by atomic hydrogen. The $H_2$ dominated "super-Earth" does not reach blow-off conditions; while the Earth-like planet can reach blow-off more or less for XUV flux values which are $\geq 100$ times that of the present Sun.

However, one should note that the modeled profiles correspond to two extreme cases. In reality one can expect that there are many upper atmospheres which will consist of $H/H_2$ mixtures which would produce profiles where the exobase levels will lie somewhere in between those shown in Figs. 7 and 8. For higher XUV fluxes the profiles shown in Fig. 7 may be closer to real scenarios, while the profiles shown in Fig. 8 will more likely represent results of lower XUV flux values.

Fig. 9 shows the corresponding temperature, density and velocity structures of the non-hydrostatic expanded upper atmosphere, which is dominated by atomic hydrogen from the lower thermosphere up to the exobase level for a higher heating efficiency value $\eta = 40$. Compared to Fig. 7 which corresponds to similar parameters but a lower heating efficiency one can see that the exobase levels expand further out and the Earth-like planet would also experience blow-off for XUV flux values which are $\geq 5$ times than that of today's solar value.

*3.3 Thermal escape rates and atmospheric loss*

By analyzing the modeled upper atmosphere response to the stellar XUV flux we find that depending on the planet's density, its skin temperature, the possible presence of IR-cooling molecules and the host stars XUV flux, the upper atmosphere experiences either high Jeans or hydrodynamic blow-off escape at high XUV fluxes, which will influence its evolution during the planet's lifetime. If the dynamically expanding bulk atmosphere does not reach the escape velocity at the exobase level and classical blow-off conditions cannot be achieved, we assume a strong Jeans-type escape. Because the classical Jeans formula is based on the isotropic Maxwellian distribution function, in such a case we have the radial velocity of the outward flowing bulk atmosphere at the exobase, and thus a distribution function which is not isotropic. Then we calculate the Jeans-type escape rate by using a shifted Maxwellian function which is modified by the radial velocity, obtained from the hydrodynamic code (e.g., Volkov *et al.*, 2011).

Recently Johnson and Volkov (2013) applied a hybrid upper atmosphere model, which combines hydrodynamic and kinetic descriptions for the study of the influence of different boundary conditions at the upper edge of the calculation domain of non-hydrostatic dynamically expanding thermospheres. They considered 3 cases: 1) matching of fluid and kinetic models, 2) a transonic assumption similar than ours, and 3) a Jeans boundary condition. For all of these cases, in spite of differences in the resulting atmospheric profiles, the total escape rates were found to be rather close to each other. This means that the escape rate is not sensitive to the type of boundary condition close at the exobase level.

If we assume a heating efficiency of 15 % we obtain for the 1 XUV case a $H_2$ outflow rate $L_{th} \approx 3.0 \times 10^{29}$ s$^{-1}$ and $\approx 5.7 \times 10^{30}$ s$^{-1}$ for atomic hydrogen which is more or less in agreement with those calculated by Tian *et al.* (2005b;see table 2 and Fig. 7, case B). By integrating the XUV heating rate over the computational domain and by taking a product of the escape rate and the potential energy of a particle one obtains the total energy deposition rate by the following equation

$$L_{th} \, G \, \frac{mM_{pl}}{R_{pl}} = \quad 5.9 \; 10^{17} \text{ erg s}^{-1} \text{ (H}_2) \quad \text{and } 6.3 \; 10^{17} \text{ erg s}^{-1} \text{ (H)}, \qquad (29)$$

where $M_{pl}$ is the planetary mass, $m$ the mass of the atmospheric species, $k$ the Boltzmann constant and $G$ the gravitational constant. The values obtained in Eq. (28) are somewhat larger than that of Tian et al. (2005b). However, it is important to note that our calculated values shown in Eq. (29) correspond exactly to the energy necessary to move the bulk atmosphere to infinity. Because atomic hydrogen will dominate over $H_2$ molecules at the exobase level and is therefore the main species which will escape from both planets we focus only on the escape rates of an upper atmosphere which is dominated by atomic hydrogen. In case a thermosphere is dominated by molecular hydrogen, the expansion of the thermosphere is less effective and the escape rate of atomic hydrogen at the exobase level would be lower compared to our values.

One can see from Fig. 3 that the irradiated part of the planet is larger than a hemisphere but does not cover the whole sphere. Therefore, we estimate the average hydrogen outflow rate to be between that for the dayside area $(2\ \pi)$ and that for the isotropic loss $(4\ \pi)$ which corresponds for the irradiated part of the planet to an average geometric factor of ~3π.

Table 3 compares the thermal H escape rates from the two test planets as a function of the assumed heating efficiency $\eta$ and XUV fluxes which are normalized to the present mean solar value (1 XUV) in 1 AU. By analyzing the results of our study according to the occurrence of hydrodynamic blow-off, we find that a hydrogen-rich, Earth-type planet even with a low $\eta$ of 15 % experiences blow-off for XUV flux values which are ≥ 10 times that of the present Sun if the planet's skin temperature at the base of the thermosphere is about 250 K. As one can see from Figs. 7 and 9, in case of the more massive "super-Earth" and $\eta = 40\%$ blow-off starts for the XUV fluxes higher than 30 and for $\eta$ of 15% for the XUV flux values which are ≥ 100 times of today's Sun.

In case the upper atmosphere experiences no blow-off conditions at the exobase level the more realistic thermal hydrogen escape rate corresponds to high thermal escape rates ($L_{thJeans-mod}$), which is based on the before explained modified shifted Maxwellian distribution function. In such cases the escape rates are slightly lower compared to the hydrodynamic outflow rate at the exobase, which contains also ballistic particles. However, if one uses the classical Jeans formula by neglecting the rapidly upward flowing atmosphere, which results in the shifted Maxwellian distribution explained above, one underestimates the escape rate to a great extent. This underestimation depends on the stellar XUV flux, the heating efficiency $\eta$ and the mean density of the planet. If blow off is not reached the more realistic escape rate in Table 3 is $L_{thJeans-mod}$ which is written in italics, but compared to the total hydrodynamic outflow rates, that includes also particle with ballistic trajectories, and the traditional Jeans loss which corresponds to Maxwellian distribution. One can see from Table 3 that the more massive "super-Earth" experiences also for higher XUV fluxes the strong Jeans-type escape $L_{thJeans-mod}$ compared to the Earth-like planet.

If we assume that our H-rich test planets orbit within the habitable zone of a solar like G star, where the XUV flux decreases over time from a factor 100 to the present solar value (1 XUV) during ~4.5 Gyr according to the power-law relationship given by Ribas *et al.* (2005), we can estimate the total thermal atmospheric hydrogen loss during that time period. Depending on the assumed heating efficiency, we obtain an estimated thermal loss of hydrogen during 4.5 Gyr for the Earth-analogue planet between ~ 5 – 11 Earth ocean ($EO_H$) amounts ($\eta$ =15 - 40%) of hydrogen and for the "super-Earth" between ~ 1.5 – 6.7 $EO_H$. ($\eta$ =15 - 40%).

We note that the total hydrogen loss estimations over the lifetime of the two test planets correspond to the XUV flux behavior of solar-like G-type stars. As it was discussed in detail in Scalo et al. (2007), lower mass M-type stars remain longer on the XUV saturation phase compared to G or F stars before the flux decreases. Therefore, if the same hydrogen-rich test planets would be located within the orbits of M dwarfs their upper atmospheres would be longer exposed to higher XUV flux values. For that reason such planets would lose more hydrogen during their lifetime. From the obtained loss rates one can see that a terrestrial planet within the orbit of the habitable zone of a Sun-like G-type star which does not lose the majority of its hydrogen-rich protoatmospheres during the earliest evolutionary stages may never get rid of it.

To get an idea what the maximum possible hydrogen escape could be, we investigate briefly under which conditions the losses may come close to the energy-limited escape rate. After the discovery of hydrogen-rich "hot Jupiters" many authors (e.g., Lammer *et al*., 2003; Vidal-Madjar *et al*., 2003; Lecavelier des Etangs *et al*., 2004; Lecavelier des Etangs, 2007; Baraffe *et al*., 2004; Erkaev *et al*., 2007; Hubbard *et al*., 2007a; 2007b; Davis and Wheatley 2009; Lammer *et al*., 2009; Leitzinger *et al*., 2011; Owen and Jackson, 2012; Lammer *et al.*, 2012a) applied the so-called energy-limited ($\eta$=100 %) escape formula (e.g., Watson *et al.*, 1981; Hunten, 1987; Lammer et al., 2003) which can be expressed as

$$L_{en} = \frac{\pi \eta R_{pl} r_{XUVeff}^{2} J_0(t)}{GM_{pl} m_H}. \tag{30}$$

Here $J_0(t)$ is the XUV flux ($\lambda = 0.1$-120 nm) outside the atmosphere as a function of stellar age at the planet's orbit location, $G$ is the gravitational constant, and $\rho_{pl}$ is the planet's mean density, and $r_{xuveff}$ is the effective radius of the XUV energy absorption (Erkav *et al*., 2012). This equation can be used to estimate the upper limit of the escape rate if a planetary atmosphere experiences hydrodynamic blow-off at its exobase level (e.g., Watson *et al.*, 1981; Zahnle *et al.*, 1988). For comparing the possible maximum loss rate estimates from Eq. (30) with the modeled thermal escape rates shown in Table 3, we present the estimates for the energy-limited escape rate with $\eta$=100 % and for the two lower $\eta$ values 40 % and 15 % in Table 4. Because Eq. (30) yields the escape rate over the whole sphere ($4\pi$) but we calculated the escape rates in Table 3 over a more realistic effective area related to $3\pi$ the escape rates given in Table 4 have been corrected by a factor of ¾.

If we compare the estimated escape rates from the energy-limited formula with those calculated by the hydrodynamic upper atmosphere model, we can see that the energy limited approach ($\eta = 100\%$ in Eq. 30) can overestimate the atmospheric escape rates, especially for lower XUV fluxes and heating efficiencies. The overestimation can be huge, especially if the upper atmosphere does not reach blow-off conditions. If we compare, for example, the escape rate from a hydrogen-rich super-Earth for the 1 and 10 XUV cases with a heating efficiency $\eta = 15$ %, due to the modified Jeans escape rate $L_{thJeans-mod}$ (Table 3: $\eta = 15\%$) with the corresponding escape rates $L_{en}$ (Table 3: $\eta = 15\%$) estimated from Eq. (30), we obtain overestimation factors of the order of ~ 36 and ~7 times. If one would not modify eq (30) by a realistic heating efficiency ($\eta = 15\%$) one would obtain an overestimation for $\eta = 100$ % compared to the $L_{thJeans-mod}$ (Table 3: $\eta = 15\%$) of the two mentioned cases of about ~ 243 and ~ 49 times. If we compare the "super-Earth" results of eq (39) for an $\eta = 15\%$ with that of the hydrodynamic model for 100 XUV and a similar $\eta$ one obtains an overestimation of ~ 2.4 times. So, our results indicate that one has to be very careful in using the energy-limited formula, even if introducing a less than 100% heating efficiency, Thus Eq. (30) may only be applied in a very restricted parameter space. This was also pointed out in the study by Lammer *et al*. (2013).

The main reason for the differences is that one has to know the right effective XUV absorption radius $r_{xuveff}$ so that the energy-limited formula that is modified by an accurate heating efficiency also yields accurate results. On the other hand without hydrodynamic simulations connected with the modeling of the stellar XUV absorption one does not obtain the right values. Therefore, if this formula is applied for XUV exposed low mass planets, including Earth-like bodies, the equation yields no accurate results if one assumes $r_{xuveff} = R_{pl}$.

For more massive planets such as hot Jupiters, Eq (30) with the assumption that $r_{xuveff} = R_{pl}$ yields escape rates which are not so different compared to hydrodynamic model results. The reason is that the large gravity of a "hot Jupiter" prevents the extreme expansion of the thermosphere compared to lower mass and size planets so that the energy deposition distributed closer to the planet's visual radius where the assumption $r_{xuveff} = r_{pl}$ accurate results. For instance, in the case of the well-studied hydrogen-rich hot gas giant HD 209458b with a skin temperature $T_0$ ~ 1350 K, and a XUV flux which is ~ 453 times higher compared to that of today's solar value, Eq. (30) leads to a thermal mass loss rate in the order of a few ~$10^{10}$ g s$^{-1}$ (e.g., Erkaev *et al*., 2007; Lammer *et al*., 2009) which is comparable with models solving similar hydrodynamic equations of mass, momentum and energy conservation as in the recent study (e.g., Yelle, 2004; Tian *et al.*, 2005b; Garcia-Muñoz, 2007; Penz *et al.*, 2008; Volkov *et al*., 2011; Koskinen *et al*., 2012).

Finally we note that hydrogen-rich planets which are not in the blow-off stage may experience high atmospheric loss rates by non-thermal processes such as stellar wind-induced ion pick-up. For understanding how efficient this non-thermal escape process could be, we use our modeled thermal escape rates as input parameters in an accompanying article by Kislyakova *et al*. (2012) who applied a coupled Direct

Simulation Monte Carlo (DSMC) stellar wind exosphere interaction model to the upper atmosphere results of the present study.

## 4 CONCLUSIONS

We studied the thermal atmospheric escape of atomic hydrogen from hydrogen-rich protoatmospheres of an Earth-like planet and a more massive "super-Earth" with $R_{pl}=2R_{Earth}$ and $M_{pl}=10M_{Earth}$ within the habitable zone of a solar type G star by applying a 1-D hydrodynamic upper atmosphere model which solves the equations of mass, momentum and energy conservation. Our results indicate that hydrogen-rich terrestrial exoplanets experience XUV-heated and hydrodynamically expanding non-hydrostatic upper atmosphere conditions during most of their lifetimes. We have found that depending on the assumed stellar XUV flux values, heating efficiencies and resulting XUV volume heating rates, hydrogen-rich terrestrial planets expand their exobase level to distances from a few $R_{pl}$ up to more than $20R_{pl}$. These expanded upper atmospheres produce huge hydrogen coronae which are most likely not protected by intrinsic magnetospheres. Earth-analogue planets reach hydrodynamic blow-off escape conditions at the exobase level inside the habitable zone for XUV flux values which are > 10 times that of today's solar value even for low heating efficiencies of 15 %.

For a higher heating efficiency of 40 %, blow-off starts at XUV fluxes which are > 5 times compared to the present Sun. Our results indicate also that hydrogen-rich more massive "super-Earths" may never reach atmospheric blow-off but experience high Jeans escape rates. For heating efficiencies which are ≥ 40 %, hydrodynamic blow-off can start also for massive "super-Earths" within the habitable zones if they are exposed to stellar XUV flux values which are > 50 times of today's solar value. In case the upper atmosphere does not experience hydrodynamic blow-off, non-thermal ion escape processes will become important. The escape rate depends on the planet's gravity and its pressure related XUV absorption distance, Only during the earliest stages in the evolution when the whole protoatmosphere was much hotter (during the first 100 Myr after the systems origin) and the stellar XUV flux was in its saturation phase, which is about 100 times that of the modern Sun the hydrogen escape rates of terrestrial planets within the habitable zone may have reached energy-limited values. After the atmosphere cooled according to the host stars luminosity at the planet's orbital location our model calculations yield total losses during 4.5 Gyr of $\sim 4.5 - 11$ EO$_H$ and $\sim 1.5 - 6.7$ EO$_H$ for the Earth-like planet and the "super-Earth", respectively, by assuming a XUV flux evolution according to Eq. (1) of Ribas *et al.* (2005). Thus, terrestrial exoplanets in orbits inside the habitable zone, which capture too much H/He nebula gas or degas dense steam atmospheres most likely keep their hydrogen envelopes and may end as sub-Neptune-type planets or low-density "super-Earths."

## 5. ACKNOWLEDGEMENTS

M. Güdel, K. G. Kislyakova, M. L. Khodachenko, and H. Lammer acknowledge the support by the FWF NFN project S116 "Pathways to Habitability: From Disks to Active Stars, Planets and Life", and the related FWF NFN subprojects, S116 604-N16 "Radiation & Wind Evolution from T Tauri Phase to ZAMS and Beyond", S116 606-N16 "Magnetospheric Electrodynamics of Exoplanets", S116607-N16 "Particle/Radiative Interactions with Upper Atmospheres of Planetary Bodies Under Extreme Stellar Conditions". K. G: Kislyakova, Yu. N. Kulikov, H. Lammer, and P. Odert thank also the Helmholtz Alliance project "Planetary Evolution and Life." P. Odert and A. Hanslmeier acknowledges also support from the FWF project P22950-N16. The authors also acknowledge support from the EU FP7 project IMPEx (No.262863) and the EUROPLANET-RI projects, JRA3/EMDAF and the Na2 science WG5. Finally, the authors thank the International Space Science Institute (ISSI) in Bern, and the ISSI team "Characterizing stellar- and exoplanetary environments". Finally, N. V. Erkaev acknowledges support by the RFBR grant No 12-05-00152-a. Finally, the authors thank referee Tian Feng, from the Tsinghua University, Beijing, China for suggestions and recommendations which helped to improve the work.

# TABLES

Table 1. Comparison of the exobase temperature $T_{exo}$, exobase distance $R_{exo}$ and the distance of the transonic point $R$s in planetary radii $R_{pl}$ for an atomic (heating efficiency $\eta = 15\%$.and 40%) molecular hydrogen thermosphere (heating efficiency $\eta = 15\%$) of an Earth-like planet, which is exposed to various stellar XUV flux values normalized to that of the present solar value.

| Earth-like: H atoms [$\eta$ = 15 %] | 1 XUV | 5 XUV | 10 XUV | 50 XUV | 100 XUV |
|---|---|---|---|---|---|
| $T_{exo}$ [K] | ~243 | ~358 | ~485 | ~1390 | ~2310 |
| $R_{exo}/R_{pl}$ | ~7.5 | ~9.5 | ~10.5 | ~16 | ~19 |
| $R_s/R_{pl}$ | ~20 | ~16 | ~12 | ~7.5 | ~6.5 |
| Earth-like: H atoms [$\eta$ = 40 %] | 1 XUV | 5 XUV | 10 XUV | 50 XUV | 100 XUV |
| $T_{exo}$ [K] | ~288 | ~575 | ~900 | ~2950 | ~4875 |
| $R_{exo}/R_{pl}$ | ~8.5 | ~10.5 | ~12 | ~18.5 | ~21.5 |
| $R_s/R_{pl}$ | ~19 | ~11 | ~9 | ~6 | ~5 |
| Earth-like: H$_2$ molecules [$\eta$ = 15 %] | 1 XUV | 5 XUV | 10 XUV | 50 XUV | 100 XUV |
| $R_{exo}/R_{pl}$ | ~5.7 | ~7 | ~7.7 | ~11 | ~12.5 |
| $T_{exo}$ [K] | ~463 | ~525 | ~625 | ~1225 | ~1875 |
| $R_s/R_{pl}$ | ~21 | ~20 | ~15 | ~9 | ~7 |

Table 2. Comparison of the exobase temperature $T_{exo}$, exobase distance $R_{exo}$ and the distance of the transonic point $R$s in planetary radii $R_{pl}$ for an atomic (heating efficiency $\eta = 15\%$.and 40%) molecular hydrogen thermosphere (heating efficiency $\eta = 15\%$) of an Earth-like planet, which is exposed to various stellar XUV flux values normalized to that of the present solar value.

| super-Earth: H atoms [$\eta$ = 15 %] | 1 XUV | 5 XUV | 10 XUV | 50 XUV | 100 XUV |
|---|---|---|---|---|---|
| $T_{exo}$ [K] | ~100 | ~625 | ~1025 | ~1575 | ~2075 |
| $R_{exo}/R_{pl}$ | ~3.5 | ~6.5 | ~8.5 | ~11 | ~13 |
| $R_s/R_{pl}$ | ~20 | ~19 | ~18 | ~15 | ~11 |
| super-Earth-like: H atoms [$\eta$ = 40 %] | 1 XUV | 5 XUV | 10 XUV | 50 XUV | 100 XUV |
| $T_{exo}$ [K] | ~500 | ~1000 | ~1250 | ~2625 | ~4050 |
| $R_{exo}/R_{pl}$ | ~7 | ~9 | ~10 | ~13.7 | ~18.5 |
| $R_s/R_{pl}$ | ~19 | ~18 | ~17 | ~11 | ~10 |
| super-Earth-like: H$_2$ molecules [$\eta$ = 15 %] | 1 XUV | 5 XUV | 10 XUV | 50 XUV | 100 XUV |
| $R_{exo}/R_{pl}$ | ~5 | ~6 | ~6.5 | ~8 | ~10 |
| $T_{exo}$ [K] | ~795 | ~1130 | ~1560 | ~2210[†] | ~2910[†] |
| $R_s/R_{pl}$ | ~22 | ~21 | ~20 | ~19 | ~17 |

[†]$T_{exo}$ values which are $\geq$ 2000 K corresponding to the thermal dissociation temperature $T_{dis}$ of H$_2$ molecules (see eq. 4).

Table 3. Calculated thermal escape rates in average over an effective area of $3\pi$ in units of s⁻¹ for H atoms and heating efficiencies $\eta$ of 15 % and 40 % for various stellar XUV flux values normalized to the XUV flux at 1 AU (habitable zone) of the present Sun for an H-rich Earth analogue planet and a "super-Earth" with the size of $2R_{\text{Earth}}$ and a mass of $10M_{\text{Earth}}$.

| H-rich Earth | 1 XUV | 5 XUV | 10 XUV | 50 XUV | 100 XUV |
|---|---|---|---|---|---|
| $L_{\text{th}}$ [s⁻¹]: $\eta$ = 15 % | ~6 × 10²⁹ | ~3 × 10³⁰ | ~ 5 × 10³⁰ | ~ 1.9 × 10³¹ | ~ 3.2 × 10³¹ |
| $L_{\text{thJeans-mod}}$ [s⁻¹] | *~1.5 × 10²⁹* | *~1.8 × 10³⁰* | | | |
| $L_{\text{thJeans}}$ [s⁻¹] | ~7.8 × 10²⁸ | ~5 × 10²⁹ | | | |
| $L_{\text{th}}$ [s⁻¹]: $\eta$ = 40 % | ~1.4 × 10³⁰ | ~ 7 × 10³⁰ | ~ 1.2 × 10³¹ | ~ 4 × 10³¹ | ~ 6 × 10³¹ |
| H-rich "super-Earth" | | | | | |
| $L_{\text{th}}$ [s⁻¹]: $\eta$ = 15 % | ~2.5 × 10²⁹ | ~1.8 × 10³⁰ | ~4 × 10³⁰ | ~ 1 × 10³¹ | ~ 2.1 × 10³¹ |
| $L_{\text{thJeans-mod}}$ [s⁻¹] | *~1.4 × 10²⁸* | *~2 × 10²⁸* | *~7 × 10²⁹* | *~ 9.2 × 10³⁰* | |
| $L_{\text{thJeans}}$ [s⁻¹] | ~1.3 × 10²⁸ | ~1.4 × 10²⁸ | ~4 × 10²⁹ | ~ 2 × 10²⁹ | |
| $L_{\text{th}}$ [s⁻¹]: $\eta$ = 40 % | ~ 1.8 × 10³⁰ | ~5.7 × 10³⁰ | ~1 × 10³¹ | ~ 2.5 × 10³¹ | ~ 4.8 × 10³¹ |
| $L_{\text{thJeans-mod}}$ [s⁻¹] | *~1.5 × 10²⁹* | *~1.6 × 10³⁰* | *~5 × 10³⁰* | | |
| $L_{\text{thJeans}}$ [s⁻¹] | ~1.4 × 10²⁹ | ~6 × 10²⁹ | ~1.3 × 10³⁰ | | |

Table 4. Maximum possible atomic hydrogen escape rates estimated with the energy-limited escape formula of eq. (28) as function of heating efficiency $\eta$ and stellar XUV flux values normalized to the XUV flux at 1 AU (habitable zone) of the present Sun for an H-rich Earth analogue planet and a "super-Earth" with the size of $2R_{\text{Earth}}$ and a mass of $10M_{\text{Earth}}$.

| H-rich Earth | 1 XUV | 5 XUV | 10 XUV | 50 XUV | 100 XUV |
|---|---|---|---|---|---|
| $L_{\text{en}}$ [s⁻¹]: $\eta$ = 15 % | ~ 6.4 × 10²⁹ | ~ 3.2 × 10³⁰ | ~ 6.4 × 10³⁰ | ~ 3.2 × 10³¹ | ~ 6.4 × 10³¹ |
| $L_{\text{en}}$ [s⁻¹]: $\eta$ = 40 % | ~ 1.7 × 10³⁰ | ~ 8.5 × 10³⁰ | ~ 1.7 × 10³¹ | ~ 8.5 × 10³¹ | ~ 1.7 × 10³² |
| $L_{\text{en}}$ [s⁻¹]: $\eta$ = 100 % | ~ 4.3 × 10³⁰ | ~ 2.1 × 10³¹ | ~ 4.3 × 10³¹ | ~ 2.1 × 10³² | ~ 4.3 × 10³² |
| H-rich "super-Earth" | | | | | |
| $L_{\text{en}}$ [s⁻¹]: $\eta$ = 15 % | ~ 5.1 × 10²⁹ | ~ 2.5 × 10³⁰ | ~ 5.1 × 10³⁰ | ~ 2.5 × 10³¹ | ~ 5.1 × 10³¹ |
| $L_{\text{en}}$ [s⁻¹]: $\eta$ = 40 % | ~ 1.3 × 10³⁰ | ~ 6.5 × 10³⁰ | ~ 1.4 × 10³¹ | ~ 6.5 × 10³¹ | ~ 1.4 × 10³² |
| $L_{\text{en}}$ [s⁻¹]: $\eta$ = 100 % | ~ 3.4 × 10³⁰ | ~ 1.7 × 10³¹ | ~ 3.4 × 10³¹ | ~ 1.7 × 10³² | ~ 3.4 × 10³² |

**FIGURE CAPTIONS**

**FIG. 1:** Illustration of the formation of dense hydrogen-rich gas envelopes and coronae around young terrestrial planets (Lammer, 2013). Panel a) Growing protoplanets can capture huge amounts of nebula-based hydrogen and He which produce dense gaseous envelopes around the rocky cores. Panel b) As soon as planetary accretion ends, depending on the interior structure and initial volatile ($H_2O$, $CO_2$, $CH_4$, $NH_3$, etc.) content of the bodies which are involved in the formation of a protoplanet, huge amounts of $H_2O$ and $CO_2$ can be released during the magma ocean solidification into the surrounding environment (e.g., Elkins-Tanton and Seager, 2008; Elkins-Tanton, 2011; Lammer, 2013). The amount of outgassed volatiles depends also on the water and carbon contents of the growing protoplanet as well as on differentiation stages and magma ocean depths. In such an outgassed steam atmosphere, the high X-ray and EUV flux of a young host star dissociates the water molecules and light H atoms populate the upper atmosphere.

**FIG. 2:** Comparison of the thermal energy flux per one steradian of the hydrodynamical flow with the thermal energy flux related only to the thermal conductivity. The dashed line shows the thermal energy flux due to hydrodynamic flow of the atmospheric particles. The dot-dashed line shows the thermal energy flux due to the thermal conductivity, which is proportional to the gradient of the temperature.

**FIG. 3**: Velocity profile obtained by our code in comparison with the analytical solution of Parker (1958): The solid curve  is the numerical solution; the "+" mark the points corresponding to the analytical solution of Parker for a Jeans escape parameter $\beta_0 = 5$.

**FIG. 4:** Illustration of the geometrical situation corresponding to our XUV flux absorption model.

**FIG. 5:** Heating rate as a function of the radial distance in the case of a XUV flux which is 20 times larger than that of today's Sun. The dotted curve is related to Eq. (21), and the dashed curve corresponds to the less complex analytical expression given by Eq. (25).

**FIG. 6:** Volume heating rate profiles up to the exobase level for 1 (long dashed lines), 5 (dashed-dotted-dotted lines), 10 (dashed-dotted lines), 50 (dashed-lines) and 100 XUV (dotted lines) cases between an Earth-like planet (6a, H atoms, $\eta$ = 15 %; 6b, $H_2$ atoms, $\eta$ = 15 %; 6c, H atoms, $\eta$ = 40 %;) and the more massive "super-Earth" (6d, H atoms, $\eta$ = 15 %; 6e, $H_2$ atoms, $\eta$ = 15 %; 6f, H atoms, $\eta$ = 40 %).

**FIG. 7:** Temperature, density and velocity profiles up to the exobase level for 1 (long dashed lines), 5 (dashed-dotted-dotted lines), 10 (dashed-dotted lines), 50 (dashed-lines) and 100 XUV (dotted lines) flux cases (Earth-like planet 7a-c, H atoms, $\eta$ = 15 %; "super-Earth" 7d-f, H atoms, $\eta$ = 15 %). The solid lines in c and f correspond to the escape velocities of the two test planets.

**FIG. 8:** Corresponding temperature, density and velocity structure of non-hydrostatic expanded upper atmospheres which are dominated by $H_2$ molecules from the lower thermosphere up to the exobase level for a heating efficiency $\eta$ = 15 % and XUV fluxes, which are 1, 5, 50 and 100 times higher compared to today's solar value for both test planets.

**FIG 9:** Temperature, density and velocity profiles up to the exobase level for 1 (long dashed lines), 5 (dashed-dotted-dotted lines), 10 (dashed-dotted lines), 50 (dashed-lines) and 100 XUV (dotted lines) flux cases (Earth-like planet 9a-c, H atoms, $\eta$ = 15 %; "super-Earth" 9d-f, H atoms, $\eta$ = 40 %). The solid lines in c and f correspond to the escape velocities of the two test planets.

**FIGURES**

**FIG 1**

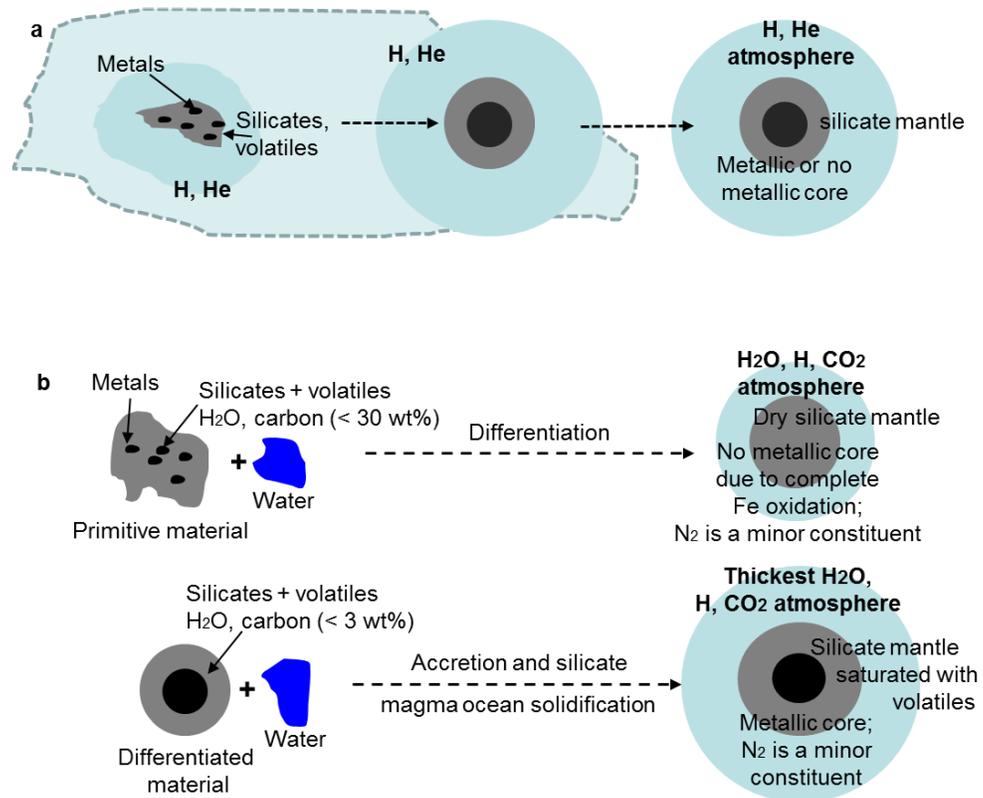

**FIG. 2**

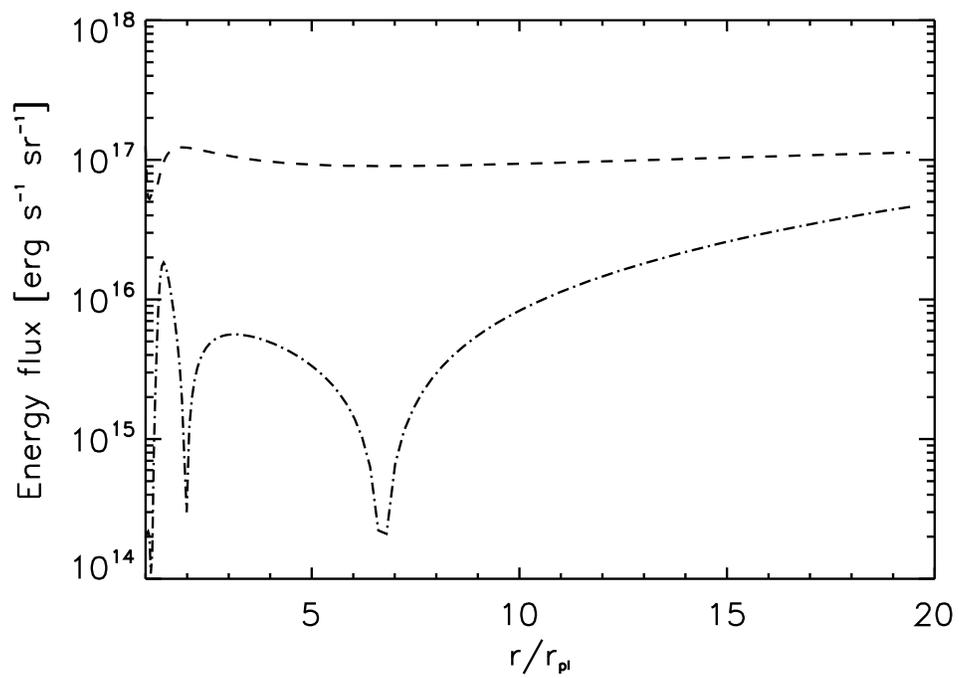

**FIG. 3**

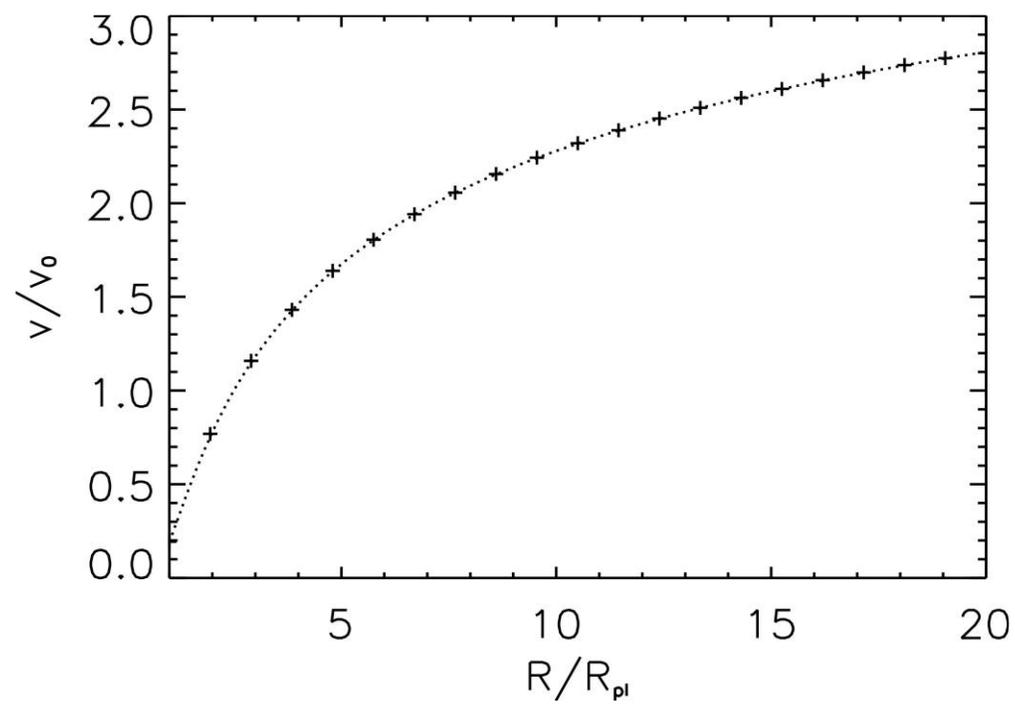

**FIG. 4**

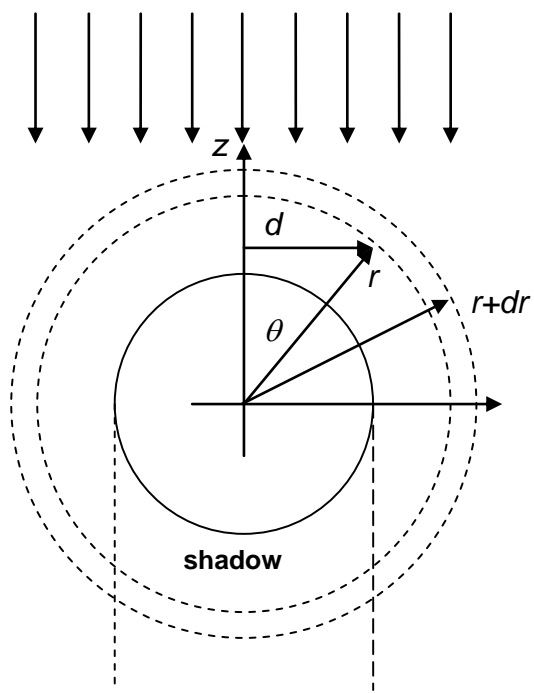

**FIG. 5**

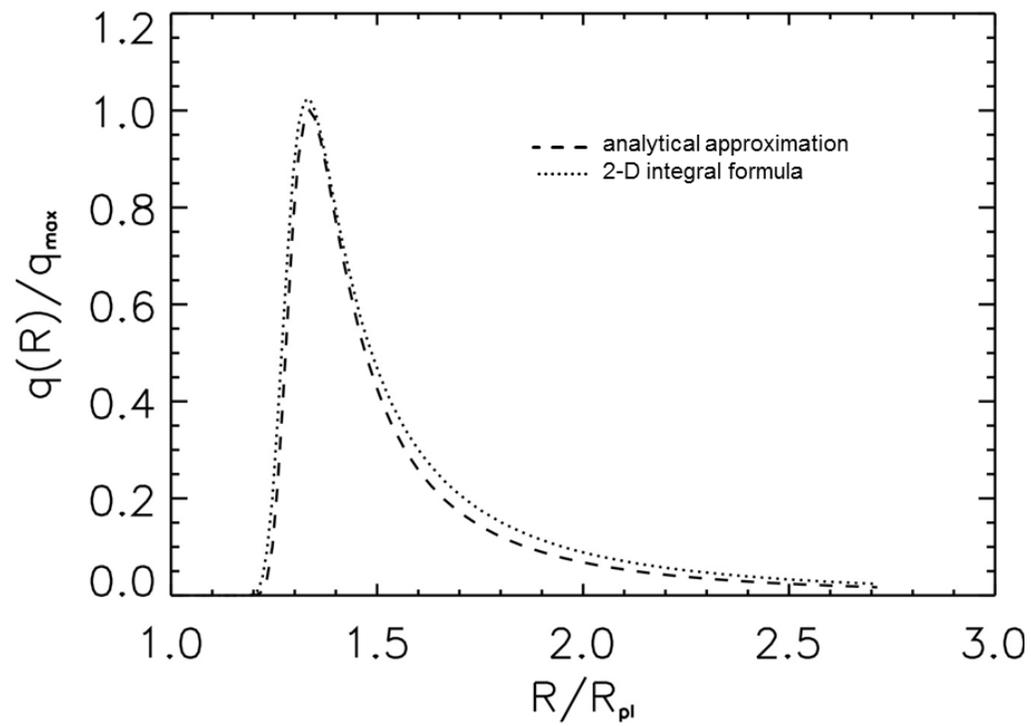

**FIG. 6**

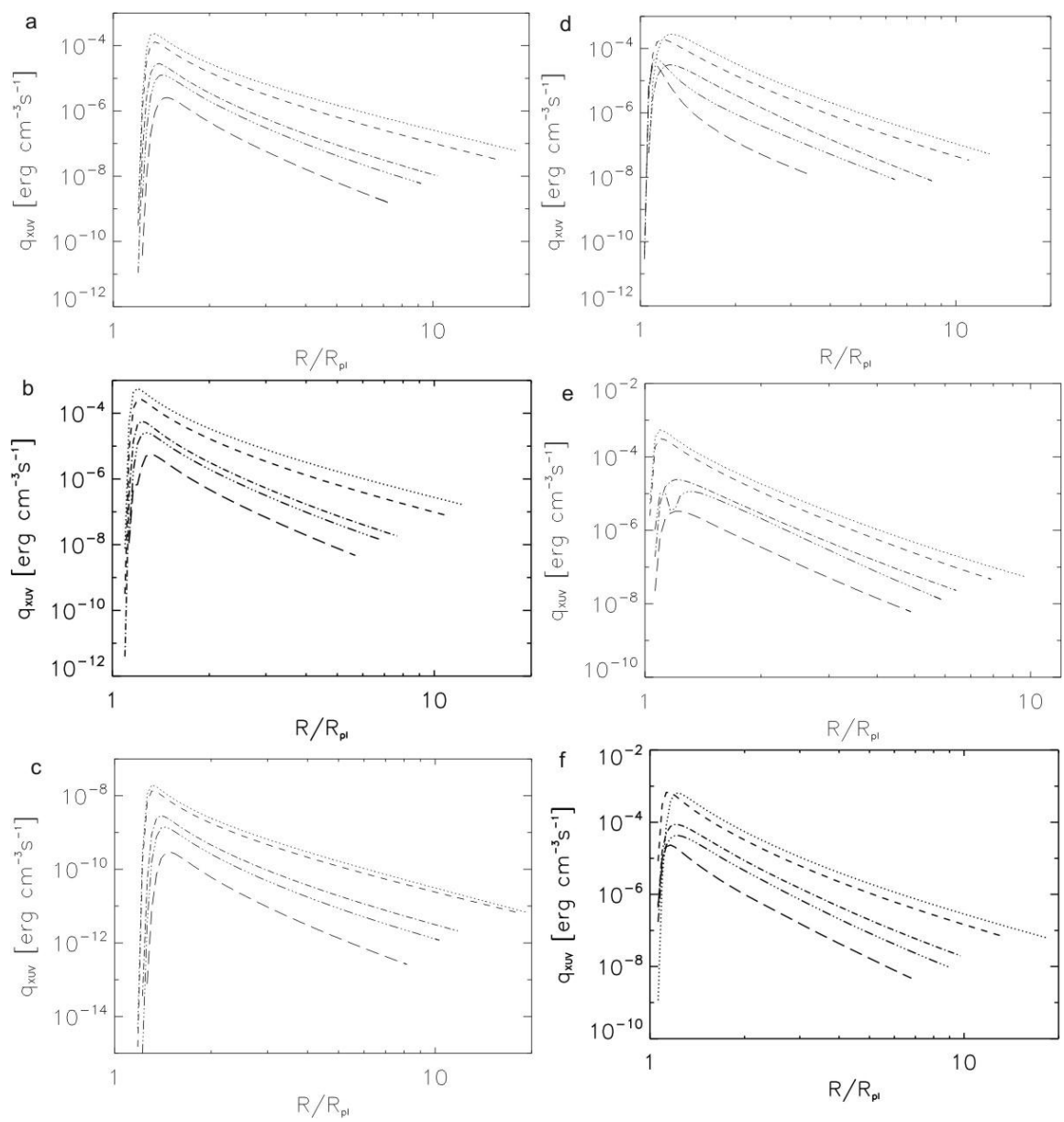

**FIG. 7**

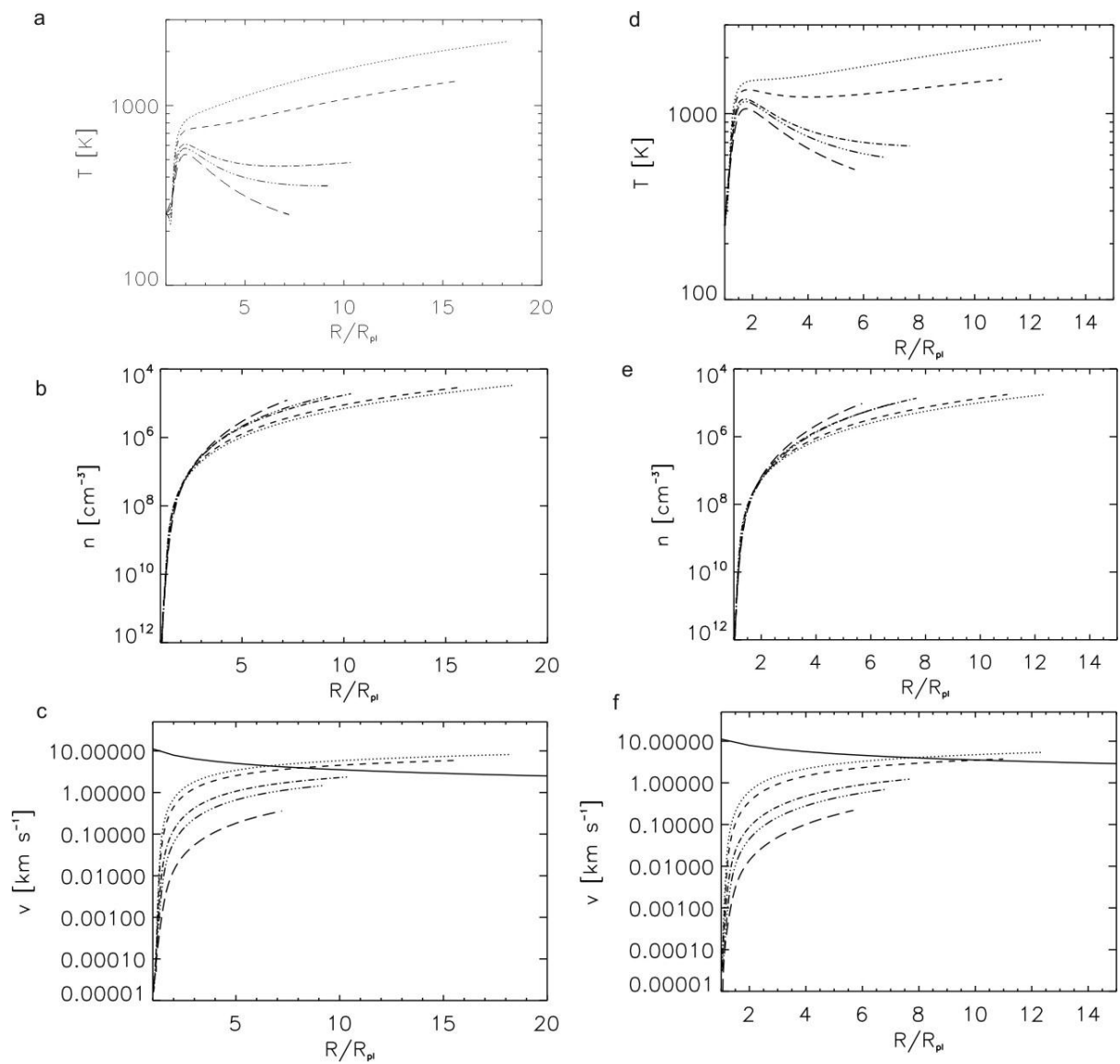

**FIG. 8**

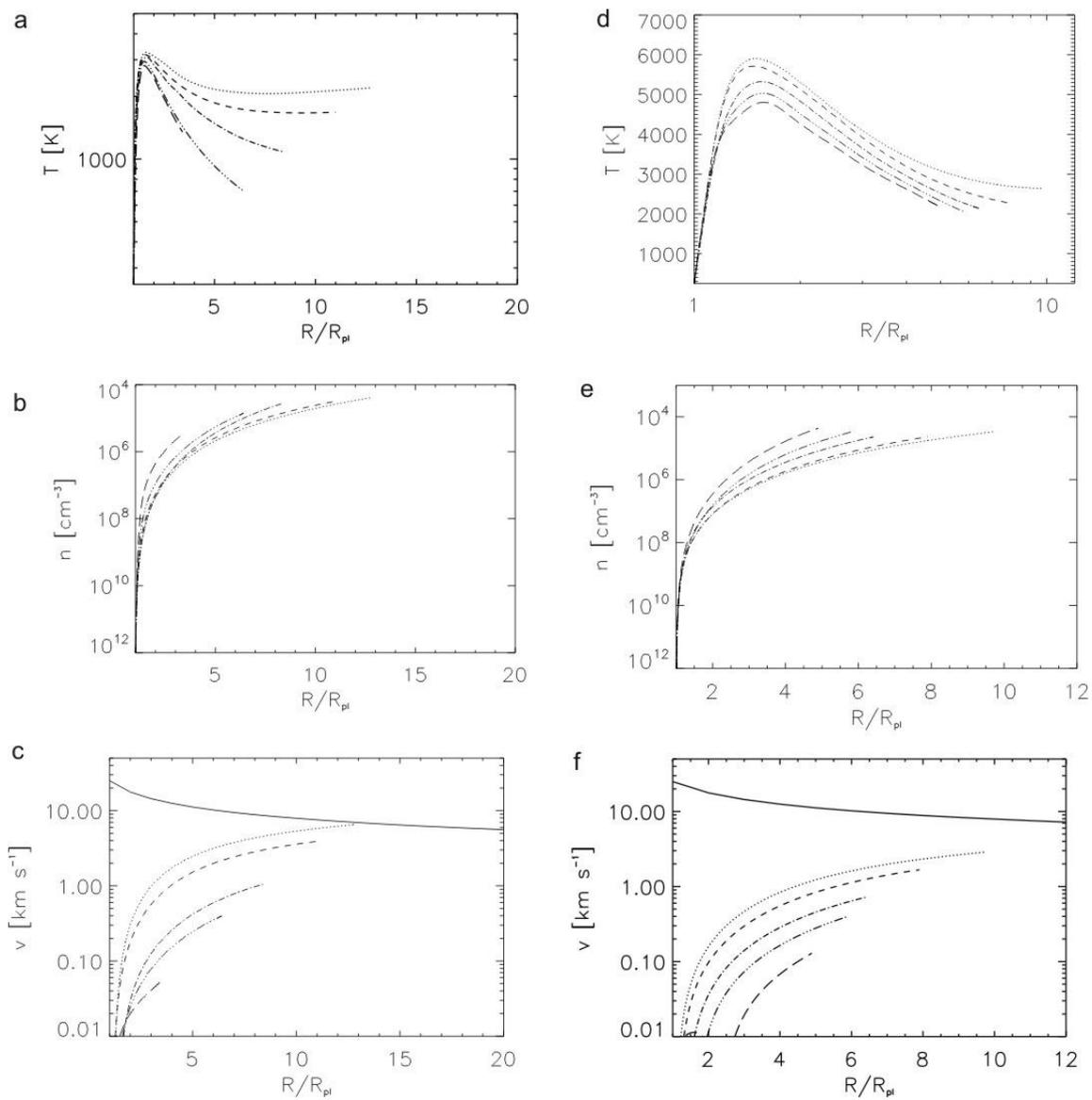

**FIG. 9**

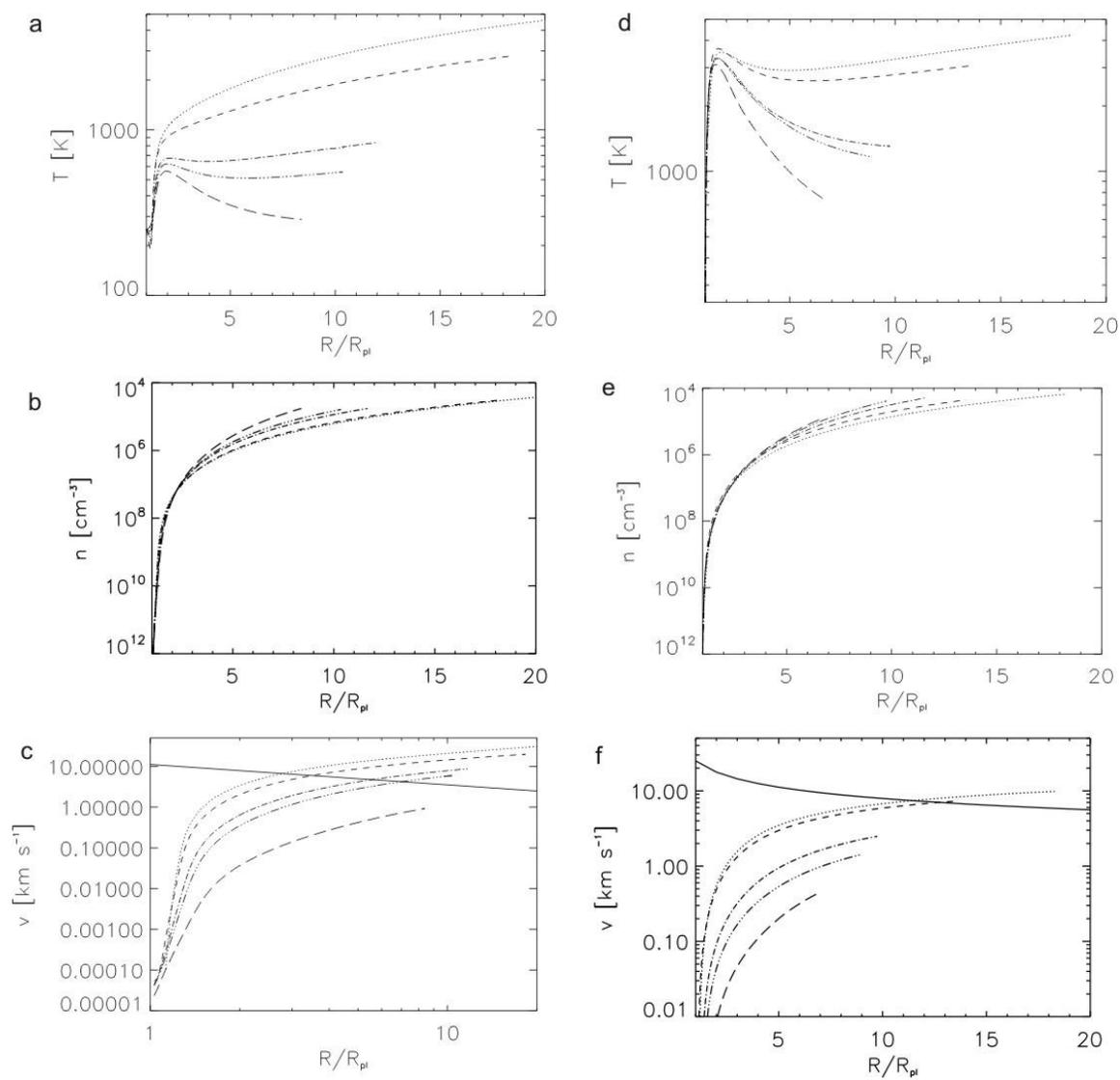